\documentclass{IEEEtran}
\IEEEoverridecommandlockouts
\usepackage{multirow}
\usepackage{subfigure}
\usepackage{stfloats}
\usepackage{soul}
\usepackage{array}
\usepackage{geometry}
\usepackage{longtable}
\usepackage{cite}
\usepackage{amsmath,amssymb,amsfonts}
\usepackage{graphicx}
\usepackage{textcomp}
\usepackage{xcolor}
\usepackage{amsthm}
\usepackage{geometry}
\usepackage{float}
\usepackage{algorithm}
\usepackage[noend]{algorithmic}
\usepackage{booktabs} 
\usepackage{array}    
\usepackage{float}    
\makeatletter
\renewcommand\subsection{\@startsection{subsection}{2}{\z@}%
                                    {-1.5ex \@plus -1ex \@minus -.2ex}%
                                    {0.5ex \@plus .2ex}%
                                    {\normalfont\normalsize\bfseries}}
\makeatother

\def\BibTeX{{\rm B\kern-.05em{\sc i\kern-.025em b}\kern-.08em
    T\kern-.1667em\lower.7ex\hbox{E}\kern-.125emX}}
\begin{document}

\title{Public Health in Disaster: Emotional Health and Life Incidents Extraction during Hurricane Harvey
}
\author{
    \begin{minipage}[t]{0.33\textwidth}
        \centering
        Thomas Hoang\\
        \textit{Department of Computer Science}\\
        \textit{Denison University}\\
        hoang$\_$t2@denison.edu
    \end{minipage}%
    \hfill
    \begin{minipage}[t]{0.33\textwidth}
        \centering
        Quynh Anh Nguyen\\
        \textit{Faculty of Information Technology}\\
        \textit{ Electric Power University}\\
        anhnq@epu.edu.vn
    \end{minipage}%
    \hfill
    \begin{minipage}[t]{0.33\textwidth}
        \centering
        Long Nguyen\\
        \textit{Department of Computer Science and Engineering}\\
        \textit{University of Louisville}\\
         l.nguyen@louisville.edu
    \end{minipage}
}

\maketitle

\begin{abstract}
Countless disasters have resulted from climate change, causing severe damage to infrastructure and the economy. These disasters have significant societal impacts, necessitating mental health services for the millions affected. To prepare for and respond effectively to such events, it is important to understand people's emotions and the life incidents they experience before and after a disaster strikes. In this case study, we collected a dataset of approximately 400,000 public tweets related to the storm. Using a BERT-based model, we predicted the emotions associated with each tweet. To efficiently identify these topics, we utilized the Latent Dirichlet Allocation (LDA) technique for topic modeling, which allowed us to bypass manual content analysis and extract meaningful patterns from the data. However, rather than stopping at topic identification like previous methods \cite{math11244910}, we further refined our analysis by integrating Graph Neural Networks (GNN) and Large Language Models (LLM). The GNN was employed to generate embeddings and construct a similarity graph of the tweets, which was then used to optimize clustering. Subsequently, we used an LLM to automatically generate descriptive names for each event cluster, offering critical insights for disaster preparedness and response strategies.
\end{abstract}
\begin{IEEEkeywords}
emotional health; climate change; large language model; graph neural network; natural language processing; social media
\end{IEEEkeywords}

\section{Introduction}
Climate change has caused many serious natural disasters around the world, like strong hurricanes, long droughts, higher temperatures, and heavy snowstorms. These extreme weather events damage buildings and the economy, affecting society deeply. Hurricanes, in particular, have become more frequent and severe. For example, Hurricane Harvey in 2017 brought massive amounts of rain to Texas and Louisiana, causing record-breaking floods. The National Hurricane Center estimated the damage at \$125 billion. Also, 738,000 people asked for help from the Federal Emergency Management Agency (FEMA), and at least 3,900 homes lost electricity \cite{amadeo2018hurricane}. The huge number of 911 calls overwhelmed emergency services, leading many people to use social media to share their problems, worries, and requests for help. Research by Cooper et al. demonstrated a strong connection between environmental conditions and emotional health through group discussions and interviews. Their study revealed that water shortages caused significant worry and fatigue among participants \cite{cooper2019environmental}. These findings were corroborated by other research, which showed that negative emotions are directly linked to immediate environmental conditions such as water shortages \cite{aihara2016household, stevenson2012water}, food shortages \cite{ojala2016young}, and environmental changes \cite{friedrich2017leading}. Hickman et al. conducted a study that highlighted the anxiety felt by many young people (aged 16-25 years) worldwide regarding climate change, with many participants expressing negative emotions towards their governments' inaction on climate issues \cite{Hickman2021survey}. To minimize bias, these studies employed various methodologies, including large surveys and group studies. Despite providing valuable insights into the impact of climate change on daily life, these studies face several challenges. Primarily, such research is often costly and time-consuming, requiring significant data collection and analysis resources. The process involves recruiting participants, organizing data collection sessions, and compensating participants, particularly in group studies. In today's world of fast technological progress and growing environmental concerns, social media platforms have become a powerful tool for investigating and understanding the different impacts of climate change. We picked this approach for a few key reasons. First, we're focusing on emotions and specific life incidents instead of just general mental health, which helps us see how environmental factors impact people's feelings during disasters. Second, we use a BERT model to predict emotions and LDA to identify life incidents, combining the power of modern NLP models and topic modeling to get accurate results. Third, we ensure our findings are reliable by automatically grouping and accurately naming the incident topics using (GNN+LLM) Graph Neural Network \cite{Kipf2016SemiSupervisedCW} \cite{Zhuang2018DualGC} and Large language Model \cite{Radford2019LanguageMA} \cite{Wei2021FinetunedLM} \cite{Brown2020LanguageMA}. Lastly, real-time social media data lets us capture public reactions and feelings immediately, giving us timely insights that are important for managing disasters and public health. While \cite{math11244910} focuses on stressors related to climate change with the use of manual topic name prediction which could be human-biased, we accurately concentrate on immediate emotional reactions and specific life incidents during disasters by leveraging the use of graph neural networks and large language model. Thus, our approach allows us to provide more detailed insights into how specific incidents affect emotional health during disasters. The collected tweets undergo an extensive data cleaning process, where URLs, special characters, and irrelevant terms are removed. We also apply stop word removal, including an expanded list to filter out common disaster-related terms that do not contribute meaningfully to our analysis. Following the cleaning process, the tweets are transformed into embeddings using a pre-trained BERT model. These embeddings are then fed into a GNN, which is trained to refine the embeddings by capturing the underlying graph structure of the data. To determine the optimal number of clusters, we employ the silhouette score, a metric that evaluates how well each tweet fits within its assigned cluster compared to other clusters. This method ensures that the tweets are accurately grouped based on their content. Once the clustering is completed, we utilize a GPT-2-based LLM to generate meaningful event names for each cluster. This step involves synthesizing the content of tweets within a cluster to predict a concise event name that encapsulates the central theme of the cluster. Our approach offers several key contributions. First, it demonstrates the effective integration of GNNs with transformer models for refining tweet embeddings, leading to more accurate clustering. Second, by using an LLM for event name generation, we move beyond traditional topic modeling, providing a more human-like interpretation of the data. Our research advances the methodological framework for disaster analysis using social media data and provides practical insights that can inform policymakers in developing comprehensive disaster management strategies that address both physical and emotional well-being.
\section{Related Studies}
In this section, we review recent studies related to addressing climate change and public health. These studies are categorized into two main scientific areas: topic modeling for public health and the use of social media for disaster relief.
\subsection{Topic modeling for public health}
Topic modeling helps find patterns and make sense of unstructured collections of documents \cite{blei2009topic}. This technique connects social and computational sciences. Topic models use probabilistic methods to uncover the hidden semantic structures of a group of texts through hierarchical Bayesian analysis. These texts can include emails, scientific papers, and newspaper articles. For example, Grassia et al. \cite{grassia2023topic} used non-negative matrix factorization (NMF) to identify main themes in newspaper articles, pinpointing topics used for propaganda. Grootendorst \cite{Grootendorst} used BERTopic to create document embeddings with pre-trained transformer-based language models, clustering these embeddings and generating topic representations with a class-based TF-IDF procedure to build neural networks. Karas et al. \cite{karas2022experiments} applied the Top2Vec model with doc2vec as the embedding model to extract topics from the subreddit "r/CysticFibrosis." Many studies use Latent Dirichlet Allocation (LDA) because it is popular and simple. For instance, Man et al. \cite{man2022evidence} used LDA to adapt an HPV transmission model to data on sexual behavior, HPV prevalence, and cervical cancer incidence. They predicted the effects of HPV vaccination on HPV and cancer incidence and the lifetime risk of cervical cancer over 100 years after vaccination. Asmundson et al. \cite{asmundson2020coronaphobia} replicated a study to examine the factor structure, reliability, and validity of the COVID-19 Incident Scales, showing how topic modeling can reveal fear and anxiety-related distress responses during pandemics. Mental health is a particular area where the importance of emotional and practical support, as well as self-disclosure, has been increasingly acknowledged. Manikonda et al. \cite{manikonda2019analysis} aimed to understand the language features, content characterization, driving factors, and types of online disinhibition seen in social media, focusing on mental health.
\subsection{Social media for disaster relief}
Social media, as explained by Kaplan, includes Internet-based applications that are built on the foundations of Web 2.0, allowing the creation and sharing of user-generated content \cite{Kaplan2018}. This term covers platforms like Reddit, Twitter, Flickr, Facebook, and YouTube, which let users communicate and share information and resources. These tools are being used more and more for disaster relief efforts. For example, Gao et al. suggested using social media to create a crowdsourcing platform for emergency services during the 2010 Haiti earthquake \cite{gao2011harnessing}. Social media can also be combined with crisis maps to help organizations find places where supplies are needed the most. A 2011 study by the American National Government looked into using social media for disaster recovery, discussing how it can be used, future possibilities, and policy considerations \cite{lindsay2011social}. Twitter, a popular social media platform, works as both a social network and a microblogging service, allowing users to post short messages called tweets. Du et al. suggested a social media-based system to analyze people's concerns, see how important they are, and track how they change over time \cite{du2019twitter}. Their study compared the flow of concerns between Twitter and news outlets during the California mountain fires. Other studies have also used social media to engage communities in water resource management \cite{nguyen2018smart}, coordinate volunteer rescue efforts \cite{yang2020coordinating}, and predict people's needs for better extreme weather planning \cite{nguyen2019forecasting}. Lu et al. visualized social media sentiment during extreme weather incidents, exploring trends in positive and negative feelings and their geographical distribution using Twitter data \cite{lu2015visualizing}. Additionally, social media can quickly assess damage from extreme weather incidents. Kryvasheyeu et al. developed a multiscale analysis of Twitter activity before, during, and after Hurricane Sandy to monitor and assess the disaster through the spatiotemporal distribution of disaster-related messages \cite{kryvasheyeu2016rapid}. 
\subsection{Graph Neural Networks}
Graph Neural Networks (GNNs) have emerged as a powerful tool for modeling relationships and dependencies in data that can be naturally represented as graphs. GNNs extend neural networks to graph-structured data, enabling the learning of representations that consider both node features and the graph structure. Kipf and Welling \cite{Kipf2016SemiSupervisedCW} introduced the concept of semi-supervised learning with GNNs, demonstrating their effectiveness in classifying nodes in a graph. This method has since been adapted to various applications, including social media analysis, where relationships between users or content can be modeled as graphs. Zhuang et al. \cite{Zhuang2018DualGC} proposed a dual graph convolutional network model, which integrates local and global graph structures to improve classification accuracy in semi-supervised settings. 
\subsection{Large Language Models for Topic Naming}
Large Language Models (LLMs) like GPT-2 have revolutionized natural language processing by enabling the generation of coherent and contextually appropriate text. Radford et al. \cite{Radford2019LanguageMA} demonstrated the capability of GPT-2 to generate text that closely mirrors human language, making it a suitable tool for creating descriptive names for clusters of events or topics. Wei et al. \cite{Wei2021FinetunedLM} further explored the adaptability of LLMs, showing that fine-tuned language models could perform well even with limited data, a common scenario in real-time social media analysis. Brown et al. \cite{Brown2020LanguageMA} introduced the concept of few-shot learning with LLMs, where the model requires minimal examples to generate relevant and specific text accurately. 
\section{Methods}
\subsection{Study Design}
\begin{figure}[H]
    \centering
    \includegraphics[width=\linewidth]{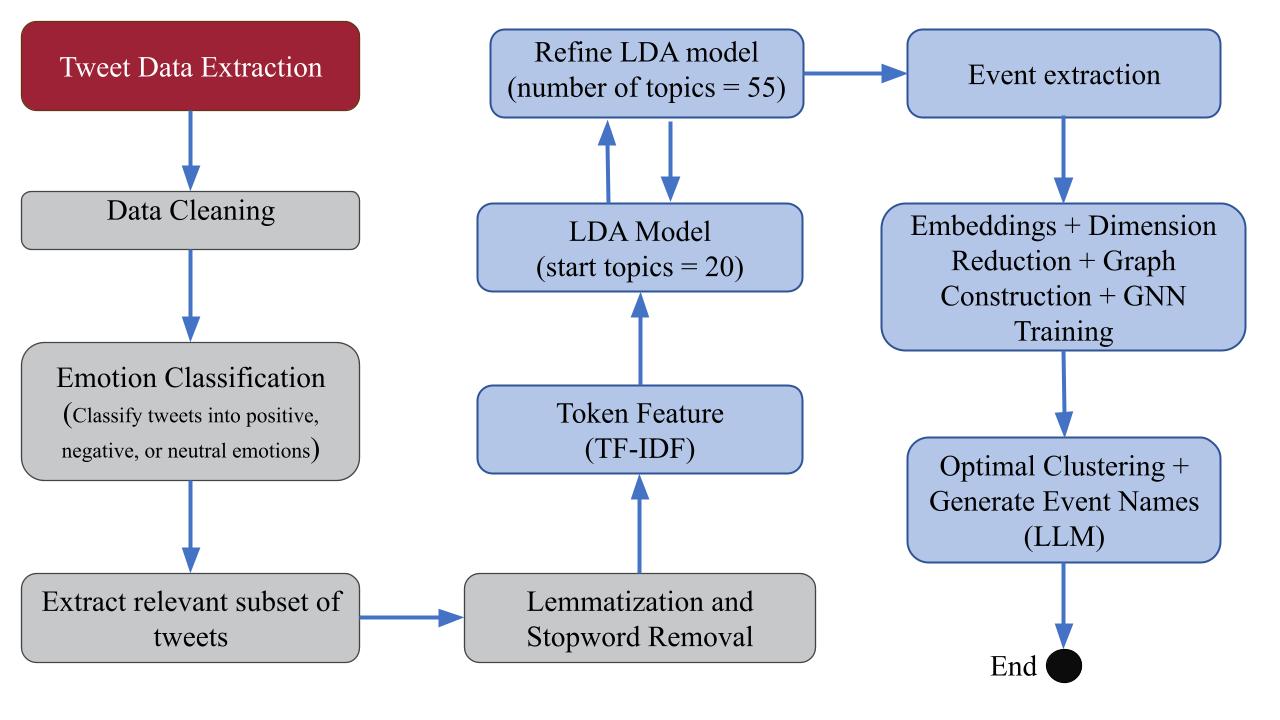}
    \caption{Overview of the research framework for climate change-related life incidents name prediction.}
    \label{fig:pipeline}
\end{figure}

We meticulously processed our collected tweet data through several stages to analyze the emotional responses to Hurricane Harvey and predict life incident names, as outlined in Figure \ref{fig:pipeline}. While in this process of cleaning data, we tried to remove emojis, hexadecimal characters, images, special characters, hyperlinks, and irrelevant words to prepare the text for analysis. Following, we tried to pass the cleaned data through an emotion classification model, which helps categorize tweets into positive, negative, or neutral sentiments. After that, we applied lemmatization to ensure that words with similar meanings but different forms (e.g., "be," "being," "been") were unified. In addition to this, we removed common English stopwords (e.g., "a," "an," "the") to eliminate non-informative words from the dataset. The text data was then transformed into token features using Term Frequency-Inverse Document Frequency (TF-IDF). With these features, we constructed an initial Latent Dirichlet Allocation (LDA) model to identify preliminary topics within the tweets. During this stage, we continuously refined our stopwords list, filtering out prevalent and unwanted tokens such as standard disaster-related terms ("hurricane," "Harvey," "storm") and location names ("Texas," "Houston," "Antonio"). Following the preliminary topic extraction, the data underwent a more rigorous processing phase, incorporating Graph Neural Network (GNN) embeddings. We performed dimension reduction on these embeddings and constructed a similarity graph, which was then used to train a GNN model. Finally, using a fine-tuned LDA model alongside the GNN-based clustering results, we employed a Large Language Model (LLM) to generate descriptive names for each predicted event group automatically. 

\subsection{Data pre-processing and feature engineering}
Our Hurricane Harvey dataset includes tweets collected from January 11, 2017, to August 29, 2017, and is publicly available on Kaggle ~\cite{harveyDataset}. The original dataset contains approximately 400,000 tweets about Hurricane Harvey. After initial filtering, we identified around 98,000 tweets expressing negative emotions. These extracted tweets then underwent data cleaning and text preprocessing to reduce redundancy and remove unwanted keywords for the topic modeling process. Specifically, we eliminated Twitter-specific characters from a defined range of Unicode characters, URLs, and hyperlinks by removing tokens containing "http." This standardization process also involved removing icons such as emojis and hex-images. Lastly, we excluded all single-character tokens from the tweets. We classify the tweets into three distinct emotion categories using BERT-based model, a Bidirectional Encoder Representations from Transformers (BERT) model with a state-of-the-art pre-built emotion detection capability.
\subsection{Emotion Prediction and life incident extraction}
\subsubsection{Text vectorization}
We employ Term Frequency-Inverse Document Frequency (TF-IDF). TF-IDF is a widely used text vectorization algorithm that creates a word frequency vector. The term frequency, inverse document frequency, and their product are computed as follows:
\begin{equation}
tf(t, d) = \dfrac{f_{t,d}}{\sum_{t'\in \; d}  {f_{t',d} }}
\end{equation}
\begin{equation}\label{eq:idf}
    idf(t, D) = \log \dfrac{ N}{1 + \lvert \left\{d \in D: t \in d \right\} \rvert}
\end{equation}
\begin{equation}
tfidf(t, d, D) = tf(t, d) \: . \: idf(t, D)
\end{equation} 

Here, \( f(t, d) \) denotes the frequency of the word \( t \) in document \( d \), and \( D \) represents the entire collection of documents. In this study, each document corresponds to a tweet. \( D \) is a corpus with a size of \( N \). To princident division by zero when \( t \) is absent in \( d \), a value of one is added to the denominator in the formula.

\subsubsection{LDA topic modeling based life incident extraction}
~\cite{chen2016plsi} demonstrates the technique of Latent Semantic Indexing (LSI) for indexing and retrieval, which helps understand the document's content by finding the relationship between words and documents. ~\cite{hofmann1999probabilistic} introduced the improvement of LSI, called probabilistic LSI (pLSI), which uses the likelihood method (e.g., Bayes method). The nature of pLSI is to help with finding the words’ models in a document where each word belongs to a specific topic. Both techniques ignore the words’ order in a document. In addition, the problem with time complexity occurs in both techniques, leading to overfitting, which Latent Dirichlet Allocation addressed well \cite{blei2003lda}. In the details of LDA, we assume we have a document (d) containing a set of words. In addition, we have a topic (z) that has several significant keywords (w). Knowing that each word can relate to many topics with various probabilities and that the amount of topics is the LDA parameter. By estimating the confidential variables ($\alpha$, $\beta$, $\theta$) by calculating the allocation in documents, LDA discovers each document's topics (Z) and the significant words of each topic. We define N as the words’ number in document $d$. Dirichlet prior parameters at the corpus level parameters are $\alpha$ and $\beta$. In addition, we choose the topic $z_{n}$ of each word from multinomial distribution $\theta$ for each word $w_{n}$. We represent as below a word $w_{n}$ from $p(w_{n} | z_{n}, \beta)$:

\begin{equation}
p(w | \alpha, \beta) = \int {p(\theta | \alpha)(\prod_{n=1}^{N} \sum_{z_n} p(z_n | \theta)p(w_n | z_n,\beta)) d{\theta}},
\end{equation}

Furthermore, we represent the probability of a corpus as below:


\begin{equation}
\prod_{d=1}^{M} \int p(\theta_d | \alpha)(\prod_{n=1}^{N_d} \sum_{z_{dn}} p(z_{dn} | \theta_d)\\p(w_{dn} | z_{dn},\beta)) d{\theta_d}
\end{equation}

\paragraph{\underline{Topics identification for optimal number:}}
In order to examine the optimal amount of topics for the LDA model, we use Umass coherence score, ~\cite{UMassscore}. This technique estimates the frequency of two words, which are $w_i$ and $w_j$: 
\begin{equation}\label{umass}
C_{UMass} = \sum_{i=2}^{N}\sum_{j=1}^{i-1}{\log\frac{P(w_i,w_j)+1}{P(w_j)}}
\end{equation}

In this equation, \( P(w_i, w_j) \) denotes the frequency with which \( w_i \) and \( w_j \) co-occur in the same document, while \( P(w_j) \) indicates the number of documents that contain the word \( w_j \). To avoid division by zero, we add a value of 1 to the denominator. The UMass coherence value is calculated as the sum of the top N pre-determined terms. Typically, \( P(w_i, w_j) + 1 \) is much smaller than \( P(w_j) \), which results in a negative UMass score. The quality of the LDA model improves as the UMass score approaches zero. However, adding more topics can increase the score, which leads to topics with very few documents. To mitigate this, we use the elbow method \cite{elbowmethod}, which helps determine the optimal number of topics by identifying the point where the rate of improvement in the UMass coherence score diminishes. After defining the topics, we manually extract the life incidents from the representative terms of each topic.

\paragraph{\underline{Life incident extraction:}}
After establishing the optimal number of topics for the LDA model, we use a Python-based LDA visualization tool to illustrate each topic and identify the key terms that influence them. This visualization helps us interpret the topics through their distinct sets of keywords. 
\begin{table}[H]
\centering
\caption{Algorithm Comparison Based on Median Silhouette Score and Purity Score.}
\resizebox{\columnwidth}{!}{
\begin{tabular}{|l|c|c|}
\hline
\textbf{Algorithm}            & \textbf{Precision Score} & \textbf{Purity Score} \\ \hline
Affinity Propagation          & 0.25                             & 0.30                 \\ \hline
Spectral Clustering           & 0.23                             & 0.25                 \\ \hline
Agglomerative Clustering      & 0.29                             & 0.25                 \\ \hline
NMF                           & 0.16                             & 0.25                 \\ \hline
Graph Neural Networks (GNN)         & 0.31                             & 0.25                 \\ \hline
\end{tabular}
}
\label{tab:comparison}
\end{table}
Based on \ref{tab:comparison}, which shows the performance among algorithm choices \cite{Frey2007ClusteringBP} \cite{Ng2001OnSC} \cite{Murtagh2011WardsHA} \cite{Lee1999LearningTP} \cite{Kipf2016SemiSupervisedCW}, we see that GNN demonstrates the best performance, with a precision score of 0.31 and a purity score of 0.25. In our case, we select that vertices in the graph represent individual terms, while edges illustrate the similarity between these terms. Thus, GNN can aggregate and propagate information across connected nodes, which leads to more accurate and contextually aware clustering and term name prediction. In detail, GNN processes the embeddings generated from the textual data, capturing both the content and the relational structure between topics. Once the topics are grouped, the LLM is used to predict descriptive names for each topic cluster. Our analysis focuses on life incidents specifically related to climate change. The GNN and LLM combination allows us to efficiently identify and name the most prominent incidents within these topics, facilitating a more detailed analysis of their impact. Thus, our method improves accuracy and enhances the extracted incidents' interpretability, making it easier to understand the specific events influencing public sentiment during disasters.
\section{Results}
\subsection{Emotion Prediction Results}
\begin{figure}[htp]
    \centering
    \includegraphics[width=\linewidth]{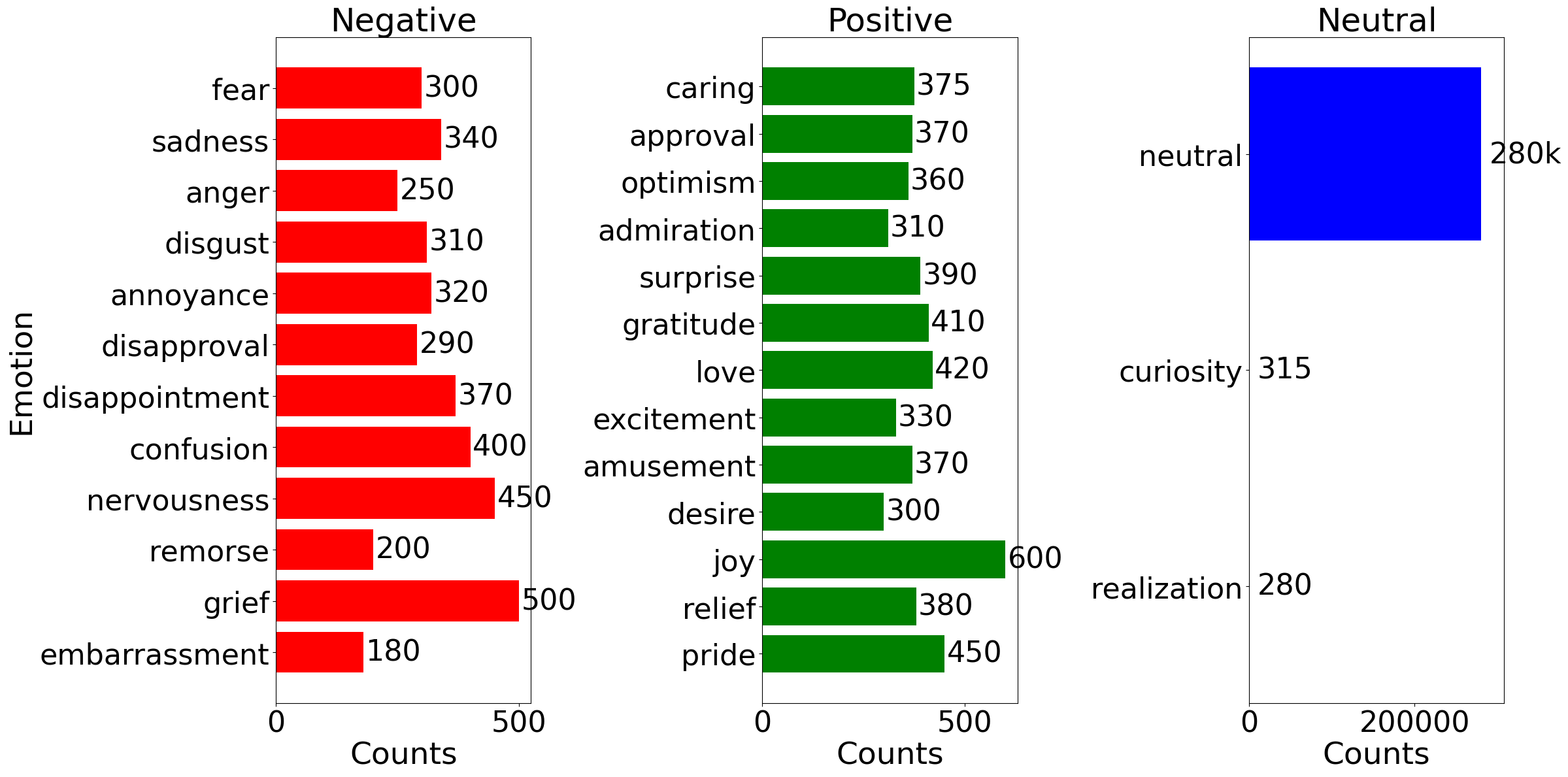}
    \caption{Distribution of predicted emotion tags in the hurricane Harvey dataset.}
    \label{fig:emotion-dist}
\end{figure}

%
We ran the algorithm using Google Collaboration, which runs on GL65 Leopard 10SCXK, an x64-based PC, on Microsoft Windows 11 Home Single Language.The emotion distribution of the tweets is illustrated in Figure \ref{fig:emotion-dist}.

\subsection{Tweets summary by emotions}
\begin{figure}[H]
\centering
\subfigure[Positive Tweets]{
\includegraphics[width=0.31\linewidth]{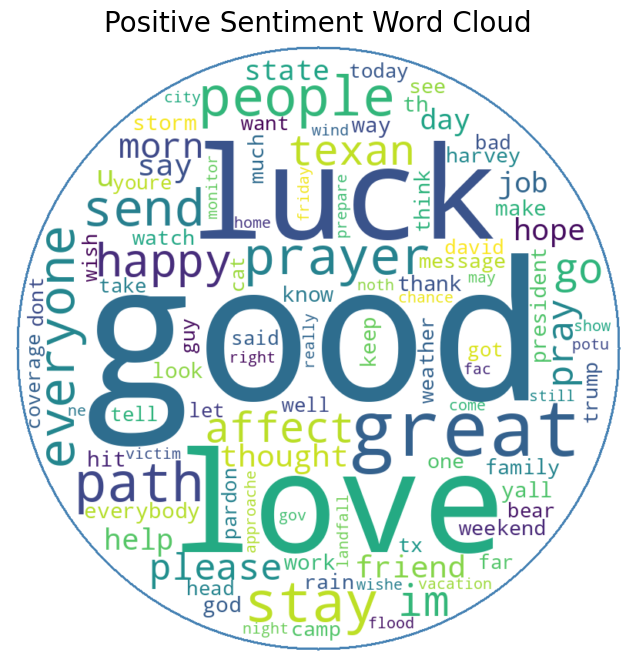}
\label{fig:positive-emotions}
}
\hspace{0.02\linewidth} 
\subfigure[Negative Tweets]{
\includegraphics[width=0.31\linewidth]{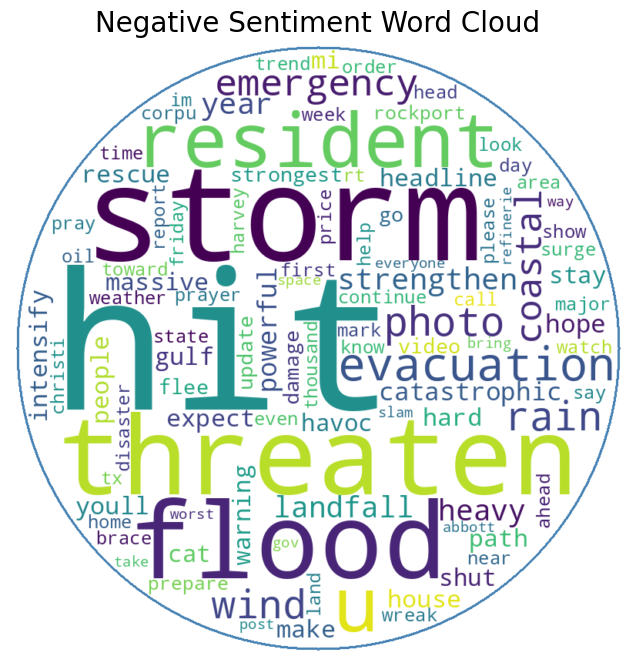}
\label{fig:negative-emotions}
}
\hspace{0.02\linewidth} 
\subfigure[Neutral Tweets]{
\includegraphics[width=0.31\linewidth]{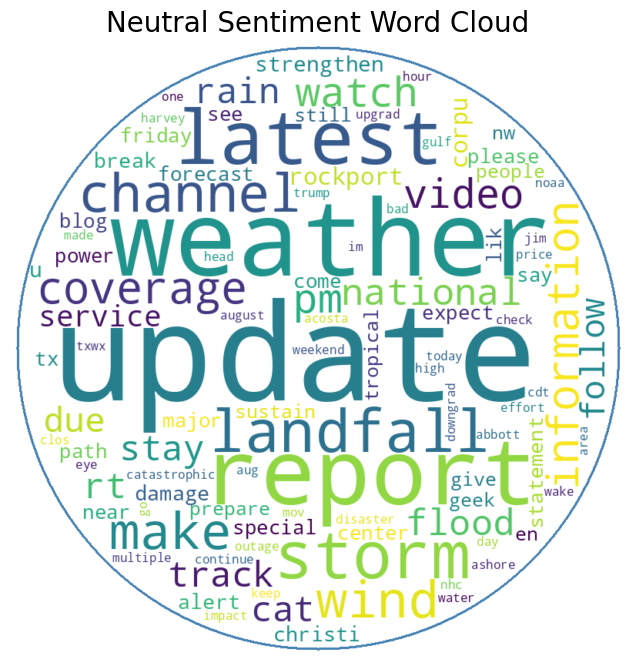}
\label{fig:neutral-emotions}
}
\caption{Overview of word clouds for Positive, Negative, and Neutral Tweets during Hurricane Harvey.}
\label{fig:tweets-summary}
\end{figure}

The positive sentiment word cloud prominently features words such as “love,” “great,” “happy,” “good,” “wonderful,” “blessed,” “safe,” and “joy.” These words reflect a general sense of optimism and positivity among Twitter users. The presence of “love” and “happy” suggests expressions of care, solidarity, and relief, possibly directed toward successful rescue operations or the safety of loved ones. These words indicate that amidst the challenges posed by the hurricane, people found moments of emotional support and happiness. The terms “great” and “good” highlight commendations and satisfaction for the effective response by emergency services or the supportive actions taken by the community. 
\begin{table}[H]
\centering
\caption{Top 20 Words for Each Sentiment Category in Hurricane Harvey Tweets based on their frequency and relevance in the tweets}
\label{tab:words_by_sentiment}
\scriptsize 
\begin{tabular}{|p{0.28\linewidth}|p{0.28\linewidth}|p{0.28\linewidth}|}
\hline
\textbf{Positive} & \textbf{Neutral} & \textbf{Negative} \\
\hline
good       & update       & hit         \\
love       & weather      & storm       \\
luck       & report       & threaten    \\
great      & latest       & flood       \\
stay       & storm        & resident    \\
people     & landfall     & u           \\
path       & channel      & evacuation  \\
prayer     & wind         & photo       \\
send       & make         & rain        \\
everyone   & information  & wind        \\
happy      & coverage     & emergency   \\
im         & watch        & coastal     \\
texan      & pm           & strengthen  \\
go         & video        & year        \\
affect     & national     & heavy       \\
safe       & hurricane    & warning     \\
wonderful  & track        & horrible    \\
blessed    & system       & damage      \\
joy        & gov          & destruction \\
support    & cnn          & disaster    \\
\hline
\end{tabular}
\end{table}
In addition, this suggests that users acknowledged and appreciated the efforts made to mitigate the disaster's impact and ensure public safety. The word “wonderful” conveys a strong sense of positivity, which might be related to successful evacuations, community support, or the resilience shown by individuals during the crisis. The appearance of “blessed” reflects a deep sense of gratitude and thankfulness, which might be in response to avoided dangers, received help, or the overall sense of being protected during the storm. 
\begin{figure}[H]
    \centering
    \includegraphics[width=\linewidth]{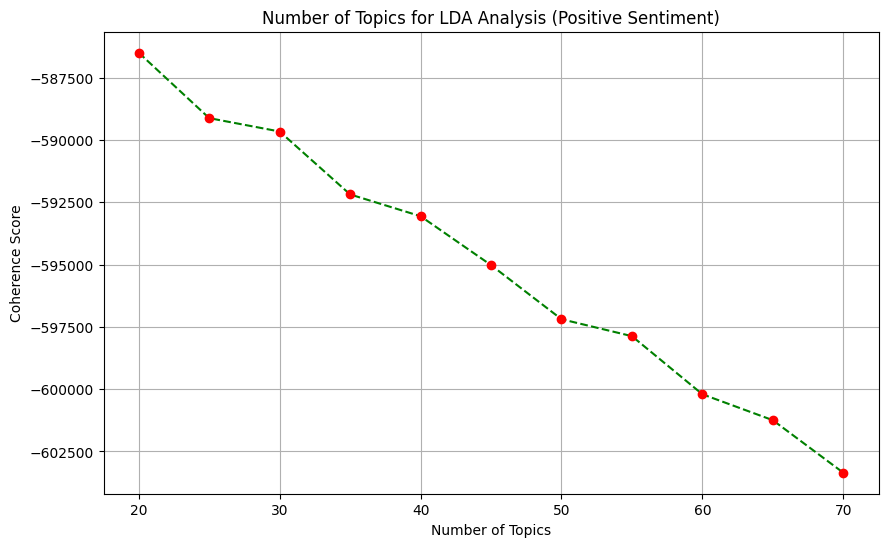}
    \caption{Number of Topics for LDA topic modeling and life incidents extraction for positive sentiments.}
    \label{fig:lda-coherrence-score}
\end{figure}
This sentiment is vital as it underscores the human aspect of the disaster response, which helps highlight moments of kindness and support that were experienced. “Safe” and “joy” further emphasize the positive outcomes and feelings of security that were felt despite the adverse conditions. These words suggest that people were able to find comfort and happiness in the safety of their surroundings or in the knowledge that their loved ones were unharmed. In general, the positive sentiment word cloud reveals a prevailing sentiment of appreciation, relief, and encouragement, reflecting the community's resilience and the successful measures taken to ensure safety and support. The positive emotions captured in these tweets highlight the human capacity to find light even in the darkest times, celebrating the small victories and the collective strength of the community.
To find the best number of topics for our Latent Dirichlet Allocation (LDA) model, we used the scikit-learn library with a learning rate of 0.7 \cite{scikit-learn}. We created several LDA models, changing the number of topics from 20 to 70 in steps of 5. Then, evaluation is done via comparing UMass coherence score \cite{UMassscore} for selection of optimal number of topics in datasets. Figure \ref{fig:lda-coherrence-score} shows an example of selection of optimal number of topics for positive sentiment. We notice that at 20 topics, the coherence score starts to decline rapidly. Thus, we chose 20 topics for the final version of our LDA model. Similarly, we get 30 topics for neutral sentiments and 55 topics for negative sentiments.
\subsection{Life incident Extraction Results}
\begin{table}[H]
  \centering
  \caption{Predicted Event Names for Positive Life Incidents Grouped by GNN and Named by LLM}
  \label{tab:positive_life_incidents}
  \scriptsize 
  \begin{tabular}{|p{6em}|p{15em}|} 
    \hline
    \textbf{Predicted Event Name} & \textbf{Representative Tweets and Terms} \\
    \hline
    \textbf{The Best of the Best} & 
    good, day, pardon, friday, happy, great, arpaio, real, im, though; \newline
    good, morn, luck, gulf, people, wish, cat, storm, love, rain; \newline
    love, job, great, good, director, handle, bug, laud, agency, help; \newline
    good, luck, love, help, victim, better, dont, deserve, near, go; \newline
    love, prayer, stay, send, path, everyone, thought, affect, good, people \\
    \hline
    \textbf{A "Good" Weather Event} & 
    good, weather, great, dog, show, food, day, side, many, bag; \newline
    great, would, love, help, could, relief, storm, change, climate, like; \newline
    good, far, im, happy, coverage, great, watch, power, get, keep; \newline
    good, luck, everybody, bear, wish, like, love, hit, bad, im; \newline
    head, good, vacation, fac, luck, great, yell, crassly, love, stay; \newline
    great, state, work, city, noth, gov, monitor, chance, federal, closely; \newline
    good, luck, great, tell, camp, david, way, president, watch, doesnt; \newline
    good, luck, path, message, people, everybody, approach, say, said, word; \newline
    god, love, great, good, hit, bless, help, thank, die, pray; \newline
    happy, love, thank, birthday, take, keep, great, ill, wait, away; \newline
    good, luck, get, corpu, go, th, look, people, like, say; \newline
    weekend, great, good, im, love, happy, let, go, cover, look; \newline
    good, great, make, landfall, go, love, impact, still, morn, wind; \newline
    pray, good, everyone, love, affect, hop, first, day, great, night \\
    \hline
  \end{tabular}
\end{table}

\section{Analysis of Optimal \( k \) Selection for Sentiment Groups}

The selection of the optimal number of clusters \( k \) for each sentiment group—negative, neutral, and positive—is informed by the silhouette score that we use to measure the quality of clustering by evaluating how similar an object is to its own cluster compared to other clusters. About positive sentiment group, for the positive sentiment group, as depicted in Figure \ref{tab:positive_life_incidents}, the silhouette score is highest at \( k = 2 \). This implies that the positive sentiment data is best categorized into two clusters, effectively capturing the key variations in positive emotions and themes expressed in the data. Having determined these optimal topic numbers and to ensure consistency and accuracy, we employ the Silhouette Score technique to automatically cluster similar groups of topics. Thanks to leveraging Graph Neural Networks (GNNs) and the use of (LLM) large language models, we can accurately generate topic names.
\begin{figure}[H]
    \centering
    \includegraphics[width=0.9\linewidth]{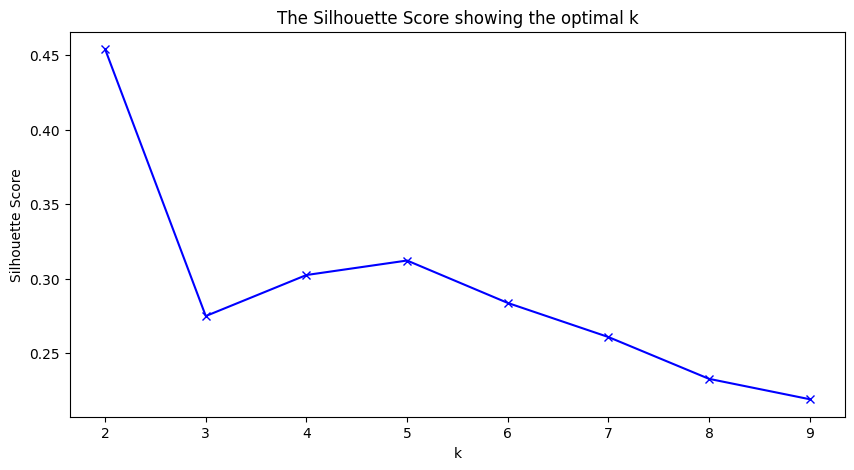}
    \caption{Comparison of Silhouette Scores for Determining Optimal k in Different Sentiment Groups}
    \label{fig:silhouette_positive}
\end{figure}

\paragraph{\underline{Life Incidents Insight Analysis}}

The extracted life incidents and their associated terms are listed in Table \ref{tab:positive_life_incidents}, representing positive sentiments. The table presents the predicted event names for life incidents grouped by a GNN-based approach and named using a large language model (LLM). Table \ref{tab:positive_life_incidents} showcases positive sentiment life incidents, which emphasize community resilience and positive interactions. The predicted event names like "The Best of the Best" and "A 'Good' Weather Event" capture the optimism and support within the community. These incidents include terms related to well-wishes, supportive actions, and positive outlooks, reflecting the community's efforts to uplift morale during challenging times.

\section{Conclusion}
Our paper presents a case study on predicting public emotions and identifying life incidents during Hurricane Harvey using social media data. We employed a Graph Neural Network (GNN) to automatically group related incidents, combined with a Large Language Model (LLM) to generate meaningful event names. Unlike previous studies that broadly examine the mental health impacts of climate change using NLP techniques, our study specifically targets emotions and life incidents during a disaster event, offering a more focused analysis of how such incidents influence public sentiment. Thus, our research will help overcome the limitations of manual extraction and enable the automated monitoring of disaster impacts on daily life and emotional health. 

\vspace{6pt}

\section{Citations}

\bibliographystyle{ACM-Reference-Format}

\end{document}